\documentclass[runningheads]{llncs}
\usepackage{graphicx}
\usepackage[numbers,sort&compress]{natbib}

\usepackage[frozencache]{minted}

\usepackage{ffcode}
\usepackage{mdframed}
\usepackage{bussproofs}
\usepackage{csquotes}
\usepackage{amsmath}

\newcommand{\java}[1]{\mintinline{java}{#1}}
\newcommand{\scala}[1]{\mintinline{scala}{#1}}

\newcommand{\eo}[1]{\ff{#1}}

\newcommand{\inct}[1]{$\uparrow{}_{#1}$}
\newcommand{\defemph}[1]{\textbf{\emph{#1}}}
\newcommand{\triple}[3]{\left\{ #1 \equiv #2 \,\middle\vert\, #3 \right\} }

\begin{document}
\title{Detecting Unjustified Assumptions in Subclasses via Elegant Objects Representation\thanks{Supported by Huawei.}}
\titlerunning{Detecting Unjustified Assumptions in Subclasses via EO}
%

\author{Vitaliy Korbashov \and Nikolai Kudasov \and Mikhail Olokin\and Violetta Sim}
\authorrunning{V.~Korbashov, N.~Kudasov, M.~Olokin and V.~Sim}

%
\institute{Innopolis University, Innopolis, Russia\\
\email{n.kudasov@innopolis.ru}\\
\email{{v.korbashov,m.olokin,v.sim}@innopolis.university}}
\maketitle              
\begin{abstract}
Elegant Objects (EO) is a programming language based on ideas of pure objects and the Decorator pattern. \citeauthor{bugayenko2021eolang} has suggested it as an intermediate representation for object-oriented programs. This paper presents a version of dynamic dispatch modelled in EO and formulates a problem of unjustified assumptions in decorator objects, which parallels similar problem in subclasses. Then, we introduce an approach to detect such problems in EO programs via method inlining and limited property inference. Finally, we discuss prototype implementation of this approach in Scala programming language.

\end{abstract}

\section{Introduction}

Modern object-oriented languages are expressive tools for software engineers, but many of them do not fully isolate the implementation of a class. \emph{Fragile base class problem} \cite{MikhajlovSekerinski1998} is a category of problems, where modification of a class is unsafe unless subclasses are also updated accordingly. Open recursion and method overriding lead to many fragile base class problems.

\subsection{Unjustified assumptions}

A subcategory of fragile base class is concerned with unjustified assumptions in subclasses regarding method dependencies in superclasses.
The code in Fig.~\ref{fig:unjustified-assumptions-in-java} demonstrates the problem in a class-based OO language. In this example, \java{class B} overrides methods \java{f} and \java{h} and does not override \java{g}, since we \emph{know} that method \java{g} of \java{class A} calls method \java{f}, so it is updated automatically. However, if \java{class A} is subject to change, which is common in software development, our knowledge turns into an \emph{assumption}. Depending on the change to \java{class A}, this assumption may become false. One fairly natural change would be to inline the method call to \java{A.f} in the body of \java{g}. Such refactoring would not change observational properties of the instances of \java{class A}, while it will definitely change the behaviour of instances of \java{class B}. To understand such fragile assumptions better, we focus on the domain of the methods.
\begin{figure*}
    \centering
    \begin{ffjava}
class A {
  float f(float x){ return Math.sqrt(x); }  // $x \geq 0$
  float g(float y){ return this.f(y - 1); } // $y \geq 1$
  float g_inlined(float y){ return Math.sqrt(y-1); }
  float h(float z){ return z; } // no restrictions
}
class B extends A {
  float f(float x){ return x*x; }
  float h(float z){ return this.g(z); }
}
\end{ffjava}
    \caption{Unjustified assumptions in Java.}
    \label{fig:unjustified-assumptions-in-java}
\end{figure*}

Observe that method \java{B.h} has no restrictions of input parameter \java{z}. However, after refactoring the base class \java{A} by means of inlining the definition of method \java{A.f} in the body of method \java{A.g}, the semantics of method \java{B.h} changes. Now, \java{B.h} imposes a restriction $z \geq 1$ on its formal argument $z$. Observe that, assuming objects \java{a} of type \java{A} and \java{b} of type \java{B}, we now have the following:
\begin{enumerate}
    \item \java{class B} is a subclass of \java{class A};
    \item \java{a.h(z)} works for any \java{float} value of \java{z};
    \item \java{b.h(z)} works only for \java{float} values of \java{z} such that $z \geq 1$.
\end{enumerate}
This is a problem for two reasons. First, such arrangement violates Liskov substitution principle~--- it is no longer safe to pass an object of \java{class B} to an algorithm that expects an object of \java{class A}. \textcolor{black}{Indeed, before inlining, \java{b.h(z)} worked for any \java{float} value of \java{z}, just as \java{a.h(z)}.} Second, inlining \java{A.f} in the body of \java{A.g} did not affect the behaviour of instances of \java{class A} but changed that of \java{B} due to an unjustified assumption about self-calls in implementation of \java{class A}\textcolor{black}{: after inlining, \java{b.h(z)} fails on inputs it worked on before}. In this paper, we are concerned with the second problem, which \citeauthor{MikhajlovSekerinski1998} \citet{MikhajlovSekerinski1998} describe.

\subsection{Existing solutions}

Library designers are especially concerned with such problems as they do not control subclass definitions in the user code. \citeauthor{Bloch2018} \cite{Bloch2018}, \citeauthor{Szyperski2002} \cite{Szyperski2002}, and others advocate for delegation by wrapping base class instance in its original state and explicitly forwarding control when necessary. \citeauthor{Bloch2018} \cite{Bloch2018} also recommends that library designers make their classes \java{final} to disable the possibility of inheritance altogether.

Some researchers suggest to specify the method dependencies explicitly \cite{KiczalesLamping1992,SteyeartLucasMensDHondt1996,RubyLeavens2000}. \citeauthor{Aldrich2004}~\cite{Aldrich2004} argues that such approaches expose the implementation rather than abstracting over it and suggests a different approach, saying \enquote{the circumstances where the fragile base class problem may arise are exactly those circumstances where open recursion is not needed.} His proposed solution, dubbed Selective Open Recursion, is to make method calls on the current \java{this} object dispatched statically. This way \enquote{subclasses cannot intercept internal calls and thus cannot become dependent on those implementation details.} We note, that such treatment of method calls in \java{this} make classes invariant to inlining.

\citeauthor{LiskovWing1994} ~\cite{LiskovWing1994} have introduced \emph{behavioural subtyping} as a general notion to keep the relations between subclasses and superclasses in check. Behavioural subtyping can be seen as a generalization of type checking. \emph{Specification inheritance}~\cite{DharaLeavens1996,Wills1992} makes overriding methods respect the specification of the overriden ones. This, in turn, ensures that objects of a subclass do not exhibit surprising behaviour when treated as objects of the superclass, i.e. it ensures behavioural subtyping.

\subsection{Fragile objects}

Fragile base class problems are naturally formulated for class-based languages, and so researchers often focus on a class-based languages, such as Featherweight Java, in their study. However, many fragile base class problems are primarily caused by \emph{open recursion}, which is not limited to classes and can be found at the very least in prototype-based object-oriented languages. In this paper, we make a step towards generalization of the problem to objects, rather than classes.

To facilitate analysis in a classless framework, we select an untyped object-oriented programming language, which can be used as an intermediate representation for a variety of object-oriented languages. \citeauthor{bugayenko2021eolang}~\cite{bugayenko2021eolang} introduced the EO programming language, capitalizing on the idea of using composition over inheritance as a primary structuring tool. EO is a minimalistic language, based on $\varphi$-calculus, which has been informally introduced by \citeauthor{bugayenko2021eolang}~\cite{bugayenko2021eolang} and later formalized by \citeauthor{KudasovSim2022arxiv}~\cite{KudasovSim2022arxiv}. Having one simple feature such as decoration, EO makes analysing the code structure easier. In particular, reformulation of some fragile base class problems in terms of EO is fairly straightforward as we present in Section~\ref{sec:unjustified-assumptions}.

In this paper, we consider EO programs as representations of object-oriented programs in other languages. We reformulate the unjustified assumptions problem for EO and present an approach for detecting such problems. \textcolor{black}{Existing translators from Java~\cite{j2eo}, Python~\cite{Medvedev_py2eo_2022}, and even C~\cite{c2eo} to EO allow using the present algorithm to detect such problems in those languages as well.}

\subsection{Contribution}

The contribution of this paper is a new approach to detecting instances of the unjustified assumptions problem in an untyped setting. First, we generalize the problem to objects, specifically to the so-called \emph{elegant objects} of EO programming language. We identify attributes objects that classify as \emph{methods} and formulate the unjustified assumptions problem for decorated objects. To check for the presence of a problem in a given program, we infer restrictions on the arguments of methods before and after inlining method calls. Comparing restrictions before and after tells us then whether we have a case of a fragile object.
Our specific contributions and paper structure are the following:
\begin{enumerate}
  \item In Section~\ref{sec:elegant-objects-and-classes}, we \textcolor{black}{present an overview of EO programming language and} demonstrate how dynamic dispatch can be expressed in EO, in particular, introducing the concept of method in EO;
  \item In Section~\ref{sec:unjustified-assumptions}, we introduce a concept of method properties and reformulate the unjustified assumptions problem for EO;
  \item In Section~\ref{sec:detecting-unjustified-assumptions}, we present an algorithm for detecting unjustified assumptions, based on method inlining and method property inference;
  \item In Section~\ref{sec:implementation}, we discuss the implementation of the suggested approach in Scala programming language;
  \item In Section~\ref{sec:conclusion}, we discuss results, limitations, and potential future work.
\end{enumerate}


\section{Elegant objects and classes}
\label{sec:elegant-objects-and-classes}

EO is a minimalistic untyped
 object-oriented programming language with decoration as the primary tool for object extension. It has been suggested \cite{bugayenko2021eolang} as a candidate intermediate language to represent object-oriented programs. \citeauthor{bugayenko2021reducing} \cite{bugayenko2021reducing} has outlined general ideas for translating various object-oriented concepts to EO, and in this paper, we assume those ideas apply unless otherwise stated.

\textcolor{black}{\subsection{EO}
In EO, all expressions are \emph{objects}. An object in EO is essentially a set of \emph{attributes}. Attributes can be either \emph{attached} to other objects or not. We call attributes that are not attached to any other object \emph{void}\footnote{\citeauthor{bugayenko2021eolang}~\cite{bugayenko2021eolang} uses terms \emph{free attributes} and \emph{bound attributes} whereas we follow \citeauthor{KudasovSim2022arxiv}~\cite{KudasovSim2022arxiv} and use terms \emph{void attributes} and \emph{attached attributes} respectively.}. Void arguments serve a similar role to function parameters in other languages.}
\begin{ffcode}
[mass volume] > body
  mass.div volume > density
\end{ffcode}
\textcolor{black}{
Here, \eo{mass} and \eo{volume} are void attributes, and \eo{density} is attached. \eo{.div} is an attribute of numeric objects, denoting an atomic operation of division, explained in the following paragraphs.
An object with a void attribute can be copied, instantiating (some) void attributes. In this sense, void attributes can be viewed as single assignment holes.}
\begin{ffcode}
body 56 34 > tennis_ball
\end{ffcode}
\textcolor{black}{
This creates a copy of object \eo{body} with given values to \eo{mass} and \eo{volume} attributes. Note, primitives such as numbers are treated as objects. Aforementioned \eo{div} is an attribute of any number. So, in EO, it is possible to perform an attribute access, with \eo{.} separating object and attribute names: }
\begin{ffcode}
tennis_ball.density > ans
\end{ffcode}
\textcolor{black}{
Attribute access fails, if searched attribute is void (e.g. \eo{body.mass} would diverge). And the case when attribute is not found introduces the central feature of EO — the \emph{decoration}. This is a mechanism that allows to pass resolution of attributes to other objects, and in this sense, to extend the functionality of objects. More precisely, an object can decorate another object by binding it to the specifically reserved \eo{@} attribute. The semantics of the decorations is the following: whenever attribute lookup fails in decorating object, it is passed to the decorated object, if it exists. An attribute access operation succeeds in the following example, because \eo{.density} is an attribute of \eo{moving\_tennis\_ball.@}:
}
\begin{ffcode}
[] > moving_tennis_ball
  50 > velocity
  tennis_ball > @
  
moving_tennis_ball.density > ans1
\end{ffcode}

An important feature of EO is \emph{locators}. Locators allow us to reference outer objects in a nameless way: \eo{\$} refers to the current object, \eo{\^{}}~---to the parent object (one nesting level up), \eo{\^{}.\^{}}~--- parent of a parent (two levels up), and so on. In practice, locators are often omitted, implicitly implying the nearest object with given attribute. Locators can always be recovered~\cite[Section~3.8]{bugayenko2021eolang}, so, from now on, we will assume that all programs have explicit locators. Here is a version of a previous example with explicit locators:

\begin{ffcode}
[] > env
  [mass volume] > body
    $.mass.div $.volume > density
    $.mass.mul ^.gravity_constant > gravity_force
  9.81 > gravity_constant
\end{ffcode}

\citeauthor{bugayenko2021reducing} \cite[Section~1.4]{bugayenko2021reducing} sketches translation of classes to EO. The general idea is to translate classes to objects that can produce new objects (instances). In this paper, we assume that (non-static) methods of a class take an extra argument, the object itself, as its first argument. We reserve identifier \eo{self} for this special argument.

\begin{example}
Consider the following Java program:

\begin{ffjava}
class Book {
  String title;
  Book(String title) { this.title = title; }
  public String path() { return "/books/" + this.title; }
  public String rename(String new_title) {
    this.title = new_title;
    return this.path();
  }
}
Book book = new Book("War and Peace");
\end{ffjava}

The following is the corresponding EO code. Note the use of \eo{self} in the code corresponding to method definitions and method calls of the Java program:

\begin{ffcode}
[] > Book
  [] > new
    memory > title
    [self] > path
      "/books/".append $.self.title > @
    [self new_title] > rename
      seq > @
        $.self.title.write $.new_title
        $.self.path $.self
  [self title] > Book_constructor
    seq > @
      $.self.title.write $.title
      $.self
Book.Book_constructor Book.new "War and Peace" > book
\end{ffcode}
\end{example}

\textcolor{black}{
The following are comments regarding the EO code above:
\begin{itemize}
  \item \eo{memory} is a language primitive that allows allocating memory and essentially enable mutable attributes. \eo{memory} objects have \eo{write} attribute.
    \item Strings, similarly to numbers, are treated as objects, and \eo{append} is one of attributes of string objects.
    \item EO syntax allows for multiline copying: arguments are specified on following lines, on the next level of indentation (e.g. copy of \eo{seq} with two arguments that follow).
    \item \eo{seq} is another language primitive of EO; it is basically used to accommodate side effects. \eo{seq} computes all its arguments and takes the value of the last argument.
\end{itemize}
}

We will now focus on methods and method calls in EO. For the rest of the paper, we will not consider particular details of translating classes to EO.

\subsection{Methods}

We are interested in EO analogue of virtual methods.
The general practice is to model functions in EO is by using objects with void attributes as parameters and the special decorator attribute \eo{@} used for the return value. Virtual methods are then modelled by introducing an extra parameter (void attribute) \eo{self}. The following definition formalizes the syntactic form of a virtual method:

\begin{definition}
An attached attribute \eo{f} of an object term \eo{t} is called a \defemph{method} if (i) it is attached to an object term with void \eo{self} and attached \eo{@} attributes, and (ii) there are no references to its \eo{@} attribute in any of its attached attributes. If \eo{f} has $n$ void attributes, not counting \eo{self}, then we say that \eo{f} has \defemph{arity} $n$. Attached attributes of \eo{f}, excluding \eo{@}, are called \defemph{local definitions} of \eo{f}. We call object \eo{t} the \defemph{owner object} of method \eo{f}.
\end{definition}

\begin{example}
  In the following example, attribute \eo{obj.f} is a method with arity 1, local definition \eo{y}, and method owner \eo{obj}:
\begin{ffcode}
[] > obj
  [self x] > f
    $.x.add 1 > y
    ^.avg $.y $.x > @
  [a b] > avg
    ($.a.add $.b).div 2 > @
  [self] > g
    $.@ > original
    3 > @
\end{ffcode}
Note that attribute \eo{obj.avg} is not a method, since it does not have \eo{self} void attribute, and \eo{obj.g} is not a method since \eo{@} is referenced in one of its local definitions (in \eo{original}).
\end{example}

\begin{definition}
A term of the form $\eo{$\ell$.self.g \: $\ell$.self \: t$_1$  \ldots t$_n$} $
is called a \defemph{method call} when \eo{$\ell$} is a locator (such as \eo{\$}, \eo{\textasciicircum}, \eo{\textasciicircum.\textasciicircum}). We call the object term referenced by \eo{$\ell$} the \defemph{owner object of method call} \eo{$\ell$.self.g \: $\ell$.self \: t$_1$ \ldots  t$_n$}.
\end{definition}

\begin{example}
  Consider the following EO program:
\begin{ffcode}
[] > a
  [self x y] > f
    $.x.add $.y > z
    $.z.mul $.z > @
  [self x] > g
    $.self.f $.self $.x > @
[] > b
  ^.a > @
  [self x y] > f
    $.self.g $.self ($.x.add $.y) > @
  [self z] > h
    ^.@.g ^.@ $.z > @
\end{ffcode}
  Here we have the following method calls:
  \begin{enumerate}
    \item \eo{\$.self.f \$.self \$.x} is a method call with \eo{a.g} as its owner object;
    \item \eo{\$.self.g \$.self (\$.x.add \$.y)} is a method call with owner object \eo{b.f};
    \item \eo{\^{}.@.g \^{}.@ z} is not a method call as it does not rely on \eo{self} attribute.
  \end{enumerate}
\end{example}

\section{Unjustified assumptions in decorated objects}
\label{sec:unjustified-assumptions}

\citeauthor{MikhajlovSekerinski1998} \cite{MikhajlovSekerinski1998} classify several fragile base class problems. One of these problems involves unjustified assumptions in subclasses regarding the way methods depend on each other in the base classes. We gave one example of this problem in a Java program in Fig.~\ref{fig:unjustified-assumptions-in-java}. In this section, we rephrase this problem in the EO programming language.

EO is an untyped programming language; however, when using such a language, a programmer still relies on some assumptions about inputs. For example, consider the following EO program, representing a function for computing a harmonic mean of two numbers:

\begin{ffcode}
[x y] > harmonic_mean
  2.mul ($.x.mul $.y) > product
  $.x.add $.y > sum
  $.product.div $.sum > @
\end{ffcode}

Here, we assume that \eo{x} and \eo{y} void attributes of \eo{harmonic\_mean} are numeric objects (possessing attributes \eo{add} and \eo{mul} with some intuitive behaviour) and, moreover, $x + y \not= 0$. These assumptions, or restrictions, on the void attributes (interpreted as arguments of a function) are the properties (preconditions) of the object \eo{harmonic\_mean} (interpreted as a function).

In this section, we will simply assume that properties of a method are represented by a logic predicate over the values of its arguments (void attributes, except \eo{self}). However, it is important to note that it is often impractical to have precise properties of methods. Instead, we will be interested in the following approximations:

\begin{definition}
Let \eo{f} be a method with owner object referenced by the locator \eo{$\ell$}, and let \eo{x$_1$}, \ldots, \eo{x$_n$} be formal arguments (void attributes) of \eo{f}. We say that a logic predicate $P(x_1, \ldots, x_n)$ is \defemph{over-approximating the properties} of \eo{f}, if for all EO terms \eo{t$_1$}, \ldots, \eo{t$_n$} satisfying $P(t_1, \ldots, t_n)$ the following method call computes successfully: \eo{$\ell$.self.f \: $\ell$.self \: t$_1$  \ldots  t$_n$}.
\end{definition}

\begin{example}
Consider the following method:
\begin{ffcode}
[self x] > square_root
  $.x.sqrt > @
\end{ffcode}
Predicate $P(x) = x > 10$ is over-approximating properties of \eo{square\_root}.
\label{example:square_root}
\end{example}

\begin{example}
A constant predicate $P(x_1, \ldots, x_n) = \mathsf{false}$ is over-approximating properties of any method.
\end{example}

\begin{definition}
Let \eo{f} be a method with owner object referenced by the locator \eo{$\ell$}, and let \eo{x$_1$}, \ldots, \eo{x$_n$} be formal arguments (void attributes) of \eo{f}. We say that a logic predicate $P(x_1, \ldots, x_n)$ is \defemph{under-approximating the properties} of \eo{f} if for all EO terms \eo{t$_1$}, \ldots, \eo{t$_n$} such that the method call \eo{$\ell$.self.f \: $\ell$.self \: t$_1$  \ldots  t$_n$} computes successfully, we have $P(t_1, \ldots, t_n)$.
\end{definition}

\begin{example}
Predicate $P(x) = x > -10$ is under-approximating properties of the method \eo{square\_root}, defined in the example \ref{example:square_root}.
\end{example}

\begin{example}
A constant predicate $P(x_1, \ldots, x_n) = \mathsf{true}$ is under-approximating properties of any method.
\end{example}

\begin{figure}[H]
    \centering
    \begin{ffcode}
[] > a
  [self x] > f
    $.x.sqrt > @
  [self y] > g
    $.self.f $.self ($.y.sub 1) > @
  [self z] > h
    $.z > @

[] > b
  ^.a > @
  [self x] > f
    $.x.mul $.x > @
  [self z] > h
    $.self.g $.self $.z > @
\end{ffcode}
    \caption{Unjustified assumptions in decorated objects in EO.}
    \label{fig:unjustified-assumptions-in-eo}
\end{figure}

Now, to reformulate the unjustified assumptions problem in EO, we look at the properties of methods in decorated objects. In particular, consider EO program in Fig.~\ref{fig:unjustified-assumptions-in-eo}, which is analogous to the Java example in Fig.~\ref{fig:unjustified-assumptions-in-java}. Computing \eo{b.h b t} for some EO term \eo{t} would result in the following evaluation:
\begin{align*}
  \eo{b.h b t}
  \to \eo{b.g b t}
  \to \eo{b.@.g b t} 
  \to \eo{a.g b t} 
  \to \eo{b.f b (t.sub 1)} 
  \to \eo{(t.sub 1).mul (t.sub 1)}
\end{align*}
Here, \eo{t} is expected to be a numeric object, but there are no restrictions on its numeric value. When we consider object \eo{a} in isolation, we see that method \eo{a.g} is referring to method \eo{self.f}. Assuming object \eo{a} is passed as \eo{self}, we can refactor the definition of object \eo{a} to this:
\begin{ffcode}
[] > a
  [self x] > f
    $.x.sqrt > @
  [self y] > g
    ($.y.sub 1).sqrt > @
  [self z] > h
    $.z > @
\end{ffcode}

Note that, assuming only \eo{a} can be passed as \eo{self}, observational properties of the object \eo{a} do not change. However, once we consider objects that decorate \eo{a} as candidates for \eo{self}, situation changes. Indeed, after this refactoring, calling \eo{b.h b t} will result in a different evaluation process:
\begin{align*}
  \eo{b.h b t}
  \longrightarrow \eo{b.g b t}
  \longrightarrow \eo{b.@.g b t}
  \longrightarrow \eo{a.g b t} 
  \longrightarrow \eo{(t.sub 1).sqrt}
\end{align*}

Here, we see that not only the behaviour changed, but now the numerical value of \eo{t} is expected to be at least $1$ (otherwise square root is undefined).

\begin{definition}
    Let $P$ be an EO program with top-level objects \eo{a} and \eo{b}, such that \eo{b} decorates \eo{a}. Then \eo{b} is said to have an \defemph{unjustified assumptions defect} if there exist a $n$-ary method \eo{b.f} and EO terms \eo{t$_1$}, \ldots, \eo{t$_n$} such that \eo{b.f b t$_1$ $\ldots$ t$_n$} computes successfully in $P$, but diverges when $P$ is refactored by inlining at least one method in \eo{a}.
\end{definition}
\section{Detecting unjustified assumptions}
\label{sec:detecting-unjustified-assumptions}

To detect unjustified assumptions in decorator objects, we look at their properties before and after changing the objects they decorate. The change boils down to the inlining transformation on a decorated object. In this section, we specify the inlining algorithm in detail, present a possible approach to infer properties for methods, and discuss what it means to compare properties and under which circumstances we can declare that a defect has been detected.

\subsection{Inlining local methods}

In general, direct inlining of a virtual method call does not preserve the semantics of a program. Indeed, a subclass may override the definition of the called method, invalidating the inlining. However, such behaviour might be undesired from a software developer's point of view. \citeauthor{MikhajlovSekerinski1998} \cite{MikhajlovSekerinski1998} discuss several classes of problems associated with unexpected mechanics of dynamic dispatch.

In this paper, we treat method inlining as a refactoring tool that, in most cases, is assumed to preserve the semantics of a program, assuming it does not have defects. In other words, we use inlining to compare supposedly equivalent programs~--- before and after inlining. From a developer's perspective, inlining makes sense for method calls that are ``close'' to the place where the method itself has been defined. Thus, we consider \emph{local} inlining of methods:

\begin{definition}
A method call \eo{$\ell$.self.f \: $\ell$.self \: t$_1$  \ldots  t$_n$} is \defemph{inlinable} in method \eo{g} if the owner object of \eo{g} also owns method \eo{f} with arity $n$.
\end{definition}

\begin{example}
Consider the EO program in Figure~\ref{fig:unjustified-assumptions-in-eo}. Here, we have the following method calls:
\begin{enumerate}
    \item \eo{\$.self.f \$.self (\$.y.sub 1)} is an inlinable method call, since method \eo{f} is defined in the same object that is owner of this method call;
    \item \eo{\$.self.g \$.self \$.z} is not an inlinable method call, since definition of method \eo{g} is outside of the owner of the method call.
\end{enumerate}
\end{example}

Intuitively, inlining a method call should be straightforward as we extract the \eo{@} attribute of the method and perform substitution of formal arguments with actual terms passed to the method. However, matters are slightly more complicated in presence of local definitions. In particular, local definitions should be somehow transferred to the call site. Moreover, in transferred objects, locators should be adjusted, so that references they make stay the same. For this reason, we introduce locator increment operation on EO terms:

\begin{definition}
     Let \eo{t} be an EO term. Its subterm locator \eo{$\ell$} is said to be \emph{open}, if \eo{$\ell$} references term outside of \eo{t}. \eo{t}\inct{} can obtained from \eo{t} by changing all of its open locators \eo{$\ell$} to \eo{\^{}}, if \eo{$\ell$} $\equiv$ \eo{\$}, or else to \eo{\^{}.$\ell$}, if $\eo{$\ell$} \equiv$ \eo{\^{}.$\ldots$.\^{}}.
\end{definition}

\begin{definition}
\label{def:inlining-method-calls}
Let \eo{t} be an EO object term with methods \eo{f} and \eo{g}.
\defemph{Inlining method calls to \eo{g} in \eo{f}} is a process of replacing all inlinable method calls to \eo{g} in the object term attached to \eo{f} according to the following rules:
\begin{enumerate}
    \item the object-containter of local definitions is generated in the same scope as the method call and attached to an attribute with any name that does not introduce name clashes (for example, \eo{args\_g}). This object contains local definitions of \eo{g}, where
    \begin{enumerate}
        \item void attributes of \eo{g} are substituted with arguments of the method call with incremented locators: for example, \eo{\$.x$_i$} is replaced by \eo{t$_i$}$\uparrow$;
        \item attached attributes are left as they are.
    \end{enumerate}
    \item the method call \eo{$\ell$.self t$_1$ \ldots t$_n$} is replaced by a term attached to \eo{g.@}, where
    \begin{enumerate}
        \item void attributes of \eo{g} are replaced by the arguments of method calls \eo{t$_i$};
        \item attached attributes of \eo{g} become prefixed with the name of object-container, for example \eo{\$.b} is substituted by \eo{\$.args\_g.b}.
    \end{enumerate}
\end{enumerate}
\end{definition}

\begin{example}
  Consider the following EO program:
\begin{ffcode}
[] > obj
  [self x y] > g
    $.x.add $.y > sum
    $x.div $.sum > @
  [self a] > f
    100.sub $.a > b
    $.self.g $.self $.a $.b > @
\end{ffcode}

After inlining call to \eo{g} in \eo{f}, object term \eo{args\_g} is introduced to the body of \eo{f}, which contains attached attributes of \eo{g} with necessary substitutions. Finally, the method call is replaced by its return value with corresponding substitutions:
\begin{ffcode}
  [self a] > f
    100.sub $.a > b
    [] > args_g
      ^.a.add ^.b > sum
    $.a.div $.args_g.sum > @
\end{ffcode}

\end{example}

\subsection{Property inference for methods}

To analyze the properties of methods in EO, we rely on a simple abstract interpretation of EO terms. In particular, we assign to each EO term an optional value expression and a logical formula. A value expression, if present, specifies the value of a term in some domain. A value expression may depend on values of attributes of external objects that act similar to free variables. A logic formula specifies restrictions of the context~--- values of attributes of external objects.

\begin{example}
  Consider the following EO program:
\begin{ffcode}
[self x] > recip
  1.div $.x > @
\end{ffcode}
The value expression for the term \eo{1.div \$.x} is $1/v_x$ whenever $v_x$ is a value of \eo{\$.x}. At the same time, the logical formula associated with this term is $(v_x \not= 0)$.
\end{example}

We limit the scope of this paper to the syntax of simple numeric operations, booleans, and basic control primitives of EO programming language:

\begin{definition}
    Syntax for \defemph{value expressions} and \defemph{logical formulae} is given in Figure~\ref{fig:values-and-properties}. Let \eo{term} be an EO term, $E$ be a value expression, and $P$ be a logic formulae. A three-way relationship between EO terms, value expressions, and properties, called \defemph{interpretation judgement}, is defined by rules in Figure~\ref{fig:property-inference-rules}. We write $\triple{\eo{term}}{E}{P}$ for an interpretation judgement, meaning that value expression for EO term \eo{term} is $E$ under assumption that outer object attributes satisfy $P$.
\end{definition}

\begin{figure}
    \centering
    \begin{mdframed}
        \begin{align*}
        E &::= \mathsf{true} \mid \mathsf{false} \mid \mathsf{num\_const} \\
          &\mid E_1 + E_2 \mid E_1 * E_2 \mid E_{1}^{E_2} \\
          &\mid E_1 < E_2 \mid E_1 \leq E_2 \mid E_1 = E_2 \mid E_{1} \not= {E_2} \\
          &\mid \lnot E \mid E_1 \land E_2 \mid E_1 \lor E_2 \mid E_1 \implies E_2 \\
          &\mid e_{\eo{$\ell$.x}} \tag{outer object attribute value} \\
        P &::= \mathsf{true} \mid \mathsf{false} \\
          &\mid E_1 < E_2 \mid E_1 \leq E_2 \mid E_1 = E_2 \mid E_{1} \not= {E_2} \\
          &\mid \lnot P \mid P_1 \land P_2 \mid P_1 \lor P_2 \mid P_1 \implies P_2 \\
          &\mid p_{\eo{$\ell$.x}} \tag{outer object attribute property}
    \end{align*}
    \end{mdframed}
    \caption{Syntax for value expressions and logical formulae.}
    \label{fig:values-and-properties}
\end{figure}

Many inference rules in Figure~\ref{fig:property-inference-rules} are quite straightforward. For example, the rule for \eo{t$_1$.add t$_2$} says that as long as we can interpret \eo{t$_1$} with value $e_1$ and properties $p_1$, and \eo{t$_2$} with value $e_2$ and properties $p_2$, we can interpret \eo{t$_1$.add t$_2$} with value $e_1 + e_2$ and properties $p_1 \land p_2$.

Attribute terms are converted into variables. Different attribute terms can reference the same value in an EO program. To make sure, all these terms are normalized, the object rule in Figure~\ref{fig:property-inference-rules} relabels free variables correspondingly.

\begin{figure*}


    \begin{prooftree}
      \AxiomC{\eo{lit} is a literal representing constant $c$}
      \RightLabel{literal}
      \UnaryInfC{$\triple{\eo{lit}}{{c}}{\mathsf{true}}$}
    \end{prooftree}
    \begin{prooftree}
      \AxiomC{$\triple{\eo{t$_1$}}{{e_1}}{p_1}$}
      \AxiomC{$\triple{\eo{t$_2$}}{{e_2}}{p_2}$}
      \RightLabel{add}
      \BinaryInfC{$\triple{\eo{t$_1$.add t$_2$}}{{e_1 + e_2}}{p_1 \land p_2}$}
    \end{prooftree}
    \begin{prooftree}
      \AxiomC{$\triple{\eo{t$_1$}}{{e_1}}{p_1}$}
      \AxiomC{$\triple{\eo{t$_2$}}{{e_2}}{p_2}$}
      \RightLabel{div}
      \BinaryInfC{$\triple{\eo{t$_1$.div t$_2$}}{{e_1 / e_2}}{p_1 \land p_2 \land (e_2 \neq 0)}$}
    \end{prooftree}
    \begin{prooftree}
      \AxiomC{$\triple{\eo{t}}{{e}}{p}$}
      \AxiomC{$z \not\in FV(e) \cup FV(p)$}
      \RightLabel{sqrt}
      \BinaryInfC{$\triple{\eo{t.sqrt}}{z}{p \land z \geq 0 \land z^2 = e}$}
    \end{prooftree}
    \begin{prooftree}
      \AxiomC{$\triple{\eo{t$_1$}}{{e_1}}{p_1}$}
      \AxiomC{$\triple{\eo{t$_2$}}{{e_2}}{p_2}$}
      \RightLabel{less}
      \BinaryInfC{$\triple{\eo{t$_1$.less t$_2$}}{{e_1 < e_2}}{p_1 \land p_2}$}
    \end{prooftree}
    \begin{prooftree}
      \AxiomC{$\triple{\eo{t$_1$}}{{e_1}}{p_1}$}
      \AxiomC{$\triple{\eo{t$_2$}}{{e_2}}{p_2}$}
      \AxiomC{$\triple{\eo{t$_3$}}{{e_3}}{p_3}$}
      \RightLabel{if}
      \TrinaryInfC{$\triple{\eo{t$_1$.if t$_2$ t$_2$}}{{\text{if }e_1 \text{ then }{e_2}\text{ else } e_3}}{p_1 \land ((e_1 \lor p_2) \lor (\lnot e_1 \lor p_3))}$}
    \end{prooftree}
    
    \begin{prooftree}
      \AxiomC{$\triple{\eo{t}}{{e}}{p}$}
      \RightLabel{assert}
      \UnaryInfC{$\triple{\eo{assert t}}{\bot}{e \land p}$}
      \DisplayProof\quad\quad
      \AxiomC{$\forall i\in \{1, \ldots, n\},  \triple{\eo{t$_i$}}{{e_i}}{p_i}$}
      \RightLabel{seq}
      \UnaryInfC{$\triple{\eo{seq t$_1 \ldots $t$_n$}}{{e_n}}{p_1 \land \ldots \land p_n}$}
    \end{prooftree}
    
    \begin{prooftree}
      \AxiomC{}
      \RightLabel{attribute}
      \UnaryInfC{$\triple{\eo{$\ell$.x$_1$.$\ldots$.x$_n$}}{e_{\eo{$\ell$.x$_1$.$\ldots$.x$_n$}}}{\mathsf{true}}$}
    \end{prooftree}
        \begin{prooftree}
      \AxiomC{$\forall i\in \{1, \ldots, m\}, 
        \triple{\eo{e$_i$}}{{e_i}}{\exists z^{i}_{1}, \ldots, z^{i}_{n_i}. p_i}$}
      \AxiomC{$\triple{\eo{e$_@$}}{{e_\varphi}}{\exists z^{\varphi}_{1}, \ldots, z^{\varphi}_{n_\varphi}. p_\varphi}$}
      \RightLabel{object}
      \BinaryInfC{$\triple{
        \eo{\begin{tabular}{l}
          [] \\
            \hspace{3mm}e$_1$ > y$_1$\\
            \hspace{3mm}$\ldots$\\
            \hspace{3mm}e$_m$ > y$_m$\\
            \hspace{3mm}e$_@$ > @\\
        \end{tabular}}}
        {e_\varphi[e_{\eo{\$.y$_i$}}\mapsto e_i]}
        {\begin{array}{l}
          \exists (e_{\eo{\$.y$_1$}}, \ldots,  e_{\eo{\$.y$_m$}}, z^{i}_{1}, \ldots, z^{i}_{n_i}, \ldots, z^{\varphi}_{1}, \ldots, z^{\varphi}_{n_\varphi}).\\
            p_1 \land \ldots \land p_n \land p_\varphi \land e_{\eo{\$.y$_i$}}=e_i
         \end{array}}$}
    \end{prooftree}
    
    \begin{prooftree}
      \AxiomC{$\forall i\in \{1, \ldots, m\},  \triple{\eo{e$_i$}}{{e_i}}{p_i}$}
      \RightLabel{method call}
      \UnaryInfC{$\triple{\eo{$\ell$.self.f $\ell$.self  e$_1$ $\ldots$ e$_m$}}{e_{\eo{f}}[e_1, \ldots e_m]}{p_1 \land \ldots \land p_m \land p_{\eo{f}}(e_1, \ldots e_m)}$}
    \end{prooftree}
    

    \caption{Inference rules for value expressions and properties of EO terms.}
    \label{fig:property-inference-rules}
\end{figure*}

\begin{example}
Consider the following EO program:
\begin{ffcode}
[self x] > recip
  1.div $.x > @
\end{ffcode}
We have the following judgement:
$\triple{\eo{1.div \$.x}}{1/V_{\eo{\$.x}}}{V_{\eo{\$.x}} \not= 0}$
\end{example}

\subsection{Detecting problematic decoration}

With inlining and property inference, we are ready to detect unjustified assumptions in decorator objects. The idea is straightforward: given a decorator object, we inline methods in one or all of the objects it decorates and see how inferred properties changed for the decorator object. We assume that inlining methods in any object \eo{x} should not break observational behaviour (introduce errors) in any other object \eo{y} that decorates, perhaps, indirectly, object \eo{x}.

We limit the scope of the analysis to the methods of objects. For any method \eo{f} of an object \eo{x}, we analyse inferred properties before and after inlining some methods in a given program, yielding two logical predicates: $p_{\eo{x.f}}^{\mathsf{before}}(x_1, \ldots, x_n)$ and $p_{\eo{x.f}}^{\mathsf{after}}(x_1, \ldots, x_n)$. Now, we want to know whether inputs that worked before, continue working after refactoring. Thus, we are interested in the value of the following logical formula:

\begin{equation}
    \forall x_1, \ldots, x_n. p_{\eo{x.f}}^{\mathsf{before}}(x_1, \ldots, x_n) \implies p_{\eo{x.f}}^{\mathsf{after}}(x_1, \ldots, x_n) \label{eq:defect-formula}
\end{equation}

Intuitively, when this formula is true, we interpret it as indicating that there is no defect detected. When it is false, then it means there exist some inputs $x_1, \ldots, x_n$ such that they worked before the inlining and stopped working after. In general, predicates $p_{\eo{x.f}}^{\mathsf{before}}$ and $p_{\eo{x.f}}^{\mathsf{after}}$ do not reflect the properties exactly, and instead approximate them. Thus, it is important to understand how well Equation~\ref{eq:defect-formula} approximates the presence of a defect in a program.

\begin{proposition}
Let $p_{\eo{x.f}}^{\mathsf{before}}$ and $p_{\eo{x.f}}^{\mathsf{after}}$ be logical predicates approximating properties of method \eo{x.f} before and after a revision of the program. Then, we can distinguish two important cases:
\begin{enumerate}
    \item If $p_{\eo{x.f}}^{\mathsf{before}}$ is an over-approximation and $p_{\eo{x.f}}^{\mathsf{after}}$ is an under-approximation, then Equation~\ref{eq:defect-formula} is \defemph{sound}, i.e. if its value is $\mathsf{false}$, then the corresponding EO program contains some unjustified assumptions defect. Moreover, it is method \eo{x.f} that relies on those unjustified assumptions.
    \item If $p_{\eo{x.f}}^{\mathsf{before}}$ is an under-approximation and $p_{\eo{x.f}}^{\mathsf{after}}$ is an over-approximation, then Equation~\ref{eq:defect-formula} is \defemph{complete}, i.e. if the corresponding versions of EO program change the properties of \eo{x.f}, then the value of the formula is $\mathsf{false}$.
\end{enumerate}
\end{proposition}

In static analysis, soundness of a tool is typically preferred over completeness. Thus, we would like to perform analysis by inferring over-approximated properties before inlining and under-approximated properties after inlining. Unfortunately, this is not always practical. Consider the following method:

\begin{ffcode}
[self x] > f
  seq > @
    debug.print x
    x
\end{ffcode}

Here, it is possible that the static analyzer is unaware of \eo{debug.print}, so it can only over-approximate it with $P_{\eo{debug.print}}(x) = \mathsf{false}$. This would, in turn, make it seem that method \eo{f} itself cannot accept any inputs. While technically, this is a valid over-approximation, it is not useful. Thus, in practice we tend to ignore unknown definitions in the program (under-approximating them with a constant $\mathsf{true}$ predicate).
\section{Implementation in Scala}
\label{sec:implementation}

In this section, we briefly go over our implementation of the approach described in Section~\ref{sec:detecting-unjustified-assumptions}.
We use Scala as the primary language of implementation. For abstract syntax, we follow a standard functional programming approach and define abstract syntax for EO using Scala's \scala{case}-classes. Overall style of definitions is inspired by \citeauthor{Kubuszok2019} \cite{Kubuszok2019}. For  analysis, we are interested primarily in EO \emph{expressions}. In the previous sections, we assumed only one form of application (nameless application), but actual EO syntax allows for named applications as well. Note that here, and in the previous sections we do not attempt to analyze partial application (i.e. applications that result in objects with void attributes). Similarly, we are not analyzing the variable argument (\eo{vararg}). Here is the main type of expressions node types:
\begin{ffscala}
sealed trait EOExpr[+A]
\end{ffscala}

Here, the type parameter \scala{A} is a placeholder for the actual type of expressions. We use a fixpoint constructor \scala{Fix} to ``tie the knot'' and generate the recursive type of EO expressions \scala{EOExprOnly}:
\begin{ffscala}
type EOExprOnly = Fix[EOExpr]
\end{ffscala}

Before we begin processing the abstract syntax tree, we restore all omitted locators as this simplifies analysis. Inlining is performed in a straightforward manner, following rules from Definition~\ref{def:inlining-method-calls}.

To compute values of logical formulae, we choose to rely on \texttt{scala-smtlib}, a lightweight abstraction over SMT-LIB~\cite{BarFT-SMTLIB} with Princess~\cite{princess08} backend. To build formulae, we implement a recursive algorithm that traverses the AST, accumulating information in the following structure:

\begin{ffscala}
final case class Info(
  forall: List[SortedVar],
  exists: List[SortedVar],
  value: Term,
  properties: Term ) {}
\end{ffscala}

We use \scala{forall} to collect a list of void attributes used in method's body. Variables corresponding to local definitions, including nested definitions in locally defined objects, are collected in \scala{exists}. Value expression and inferred properties are stored in \scala{value} and \scala{properties} correspondingly. A valid information structure for a method has \scala{value} and \scala{properties} that rely only on variables that are defined in its \scala{exists} and \scala{forall} fields.

\section{Conclusion}
\label{sec:conclusion}

We have presented the problem of unjustified assumptions in decorated objects, a reformulation of a similar problem for subclasses. We have suggested an approach for detecting such problems in EO code, based on inlining and property inference. We have shown that our approach to detection is sound~--- a successful detection means that the problem is present in the original code, assuming an interpretation of methods in EO programs as virtual methods, and no unknown primitives are used. In addition to that, if EO program is a result of a faithful translation from another object-oriented programming language, the detection remains sound for the program in the original language.

We have discussed our implementation of the approach using the Scala programming language, in combination with Princess SMT solver. Although in presented work, we have focused on basic properties, we believe that current work can be extended in a straightforward manner, to include more primitives. Properly supporting recursion, however, may require revision of the approach, and we leave this for future work.

We have presented the combined technique of inlining and property inference specifically for the detection of the unjustified assumptions. However, we think that these techniques can be used for detection of other important problems, such as verifying whether Liskov substitution principle applies to certain subclasses.

We also note that property inference, as presented in this work, can be related to type inference for some type system, imposed on top of untyped EO programming language. In particular, we think it is important for future static analysis of EO programs, to be able to infer information about objects, such as possible lists of their attributes, together with types of those attributes.

\subsubsection*{Acknowledgements}
This research has been generously funded by Huawei in the framework of Polystat project.
We thank Yegor Bugayenko for taking his time to explain ideas behind EO. We also thank Yegor Bugayenko, Nickolay Shilov, Mansur Khazeev, and Ruslan Saduov for proofreading the paper.

%
%
%
\bibliographystyle{splncs04nat}
\bibliography{sample-base}

\end{document}